\documentclass[floatfix,showkeys,showpacs,twocolumn,amsmath,amssymb,epsf,aps]{revtex4}
\usepackage{dcolumn}
\usepackage{epsfig}

\usepackage{graphicx}
\usepackage{longtable}
\usepackage{dcolumn}
\usepackage{bm}

\def\etal{{\em et al. }}

\def\cuo #1{CuO${_#1}$}

\def\om #1{$\omega{_#1}$}
\def\cmin {cm$^{-1}$}
\def\fn{\footnotemark}
\def\YY{{\bf Y }}
\def\BB{{\bf B }}
\def\AAA{{\bf A }}
\def\CC{{\bf C }}
\def\DD{{\bf D }}

\def\GG{{\bf G }}
\def\FF{{\bf F }}

\begin{document}
\preprint{USC/002}

\title{ Molecular Structures and Vibrations of neutral and anionic CuO$_x$ (x=$1-3$,6)  clusters}

\author{Tunna Baruah$^{1,3}$, Rajendra R. Zope$^2$, and Mark R. Pederson$^3$}
\affiliation{$^1$Department of Physics, Georgetown University, Washington DC}
\affiliation{$^2$School of Computational Sciences, George Mason University, Fairfax,
VA 22030}
\affiliation{$^3$Center for Computational Materials Science, Naval Research 
Laboratory, Washington DC 20375-5345}

 \email{baruah@dave.nrl.navy.mil}
 \email{rzope@scs.gmu.edu}
 \email{pederson@dave.nrl.navy.mil}

\date{\today}

\pacs{ }


\begin{abstract}

    We report equilibrium geometric structures of  \cuo{2},  \cuo{3}, \cuo{6}, and CuO$_6^{-1}$ 
clusters obtained by an all-electron linear combination of atomic orbitals scheme within the 
density functional theory with generalized gradient approximation to describe the exchange-correlation 
effects.  The vibrational stability of all clusters is examined on the basis of the vibrational frequencies.
A structure with C$_s$ symmetry is found to be the lowest energy structure for \cuo{2} while
a \YY shaped structure with C$_{2v}$ symmetry is the most stable structure for \cuo{3}. For the larger 
\cuo{6} and CuO$_6^{-1}$ clusters, several competitive structures exist with
structures containing ozonide units being higher in energy than those with O$_2$ units.  The infra-red and 
Raman spectra are calculated for the stable optimal geometries.

\end{abstract}

\maketitle

\section{\label{sec:intro}Introduction\protect\\ }

     The copper-oxygen complexes have received significant attention
 because of their role in various processes such as catalytic oxidation,  
applications in the bio-inorganic chemistry, and corrosion to name a few.
Numerous studies, both theoretical as well as 
experimental, have been reported to elucidate the chemical bonding between
copper  and oxygen in such complexes.  Most of these studies were confined to small sizes 
of \cuo{x} systems such as CuO and CuO$_2$ molecules \cite{WDW97,WWDL96, PM00,PMC00,PM02}. 
These earlier studies have been recently extended to larger systems
by Wang  and coworkers  \cite{WWDL96, WDW97} who  reported photoelectron spectra (PES)
studies  of \cuo{x} $(x=1-6)$ species. They found that the Cu/O
complexes containing either an odd or even number of oxygen atoms 
had similarities among themselves with respect to their PES. 
They also found several new copper-oxide molecules including \cuo{6} and proposed some 
structures for these molecules on the basis of their PES study. 
 Massobrio and coworkers\cite{PM02,PM00,PMC00} subsequently performed a systematic search
 for the lowest energy
 structure for \cuo{6}. In their density functional calculations with ultrasoft pseudopotentials
 they found the lowest energy structure of \cuo{6} to possess ozonide Cu(O$_3$)$_2$ units
 contrary to the prediction in Ref.\onlinecite{WDW97} of an oxide structure Cu(O$_2$)$_3$ on 
the basis of photo-electron spectra. They also found
 that the structures containing the Cu(O$_2$)$_3$ units to be roughly within 0.2 eV. 
Cu is known to form OCuO or Cu(O$_3$) upon reaction with O$_2$ only under UV radiation
at low temperature \cite{OMG83}.
The Cu atom at room temperature can form only Cu(O$_2$) complexes with O$_2$ because the
O atom transfer reaction is endothermic. 

 The present work is aimed at providing a detailed picture of the stationary points on
 the potential energy surface of the  \cuo{2}, \cuo{3}, \cuo{6} and anionic
 \cuo{6} clusters.  For this purpose, we perform a systematic search for possible
 candidate structures for the ground state geometries of these clusters,
 followed by calculation of the vibrational spectra  to identify the nature of
 stationary points. As the energy differences between the isomers 
are small, we employ an all-electron treatment thereby eliminating any question 
due to the ultrasoft pseudopotential approximation. The present work complements the previous
investigation on these systems by Pouillon and coworkers \cite{PMC00,PM00,PM02}. 
Additionally, we have also calculated vertical detachment energies of the \cuo{6} 
isomers and also the infrared absorption and Raman scattering spectra of the
lowest energy \cuo{6} isomers. These may be useful for comparison with
experimental measurements.

 In the following section the computational procedures
employed in this work are briefly described followed by the results
and discussion in the subsequent section.

\section{\label{sec:method}Details of computational procedure\protect\\ }
  
  The calculations have been performed within the Kohn-Sham formulation 
of the density functional theory with the generalized gradient approximation (GGA)
to describe the exchange-correlation effects \cite{DFTPY}. We have used the PBE functional
by Perdew and coworkers \cite{PBE}.  Structural optimizations of all the clusters
were performed by starting from a number of possible candidate structures,
including those already reported in the literature \cite{PM00,PM02}. The optimization 
was performed
using the conjugate-gradient as well as the LBFGS method.  For the  case of the anionic \cuo{6}
systems, the optimized neutral geometries are used as the starting geometries.
All calculations have been performed using the 
Naval Research Laboratory Molecular Orbital Library (NRLMOL) package \cite{NRLMOL}. 
This suite of programs employs a Gaussian basis set  where
the exponentials of the Gaussian basis are optimized for each atom \cite{Porezag99}. The integrations
are carried out using a grid based method with  a variational mesh.
Further, the  magnetic moment is optimized during the self-consistency cycle
from electron counting considerations.  The geometries were optimized until
the forces were less than 0.001 a.u..
The nature of stationary points (optimized geometries) on the potential energy surface 
was examined by computing the vibrational frequencies.
These were calculated by displacing the atoms from their equilibrium positions
by $\pm$0.07 a.u. along each Cartesian direction and constructing the dynamical
matrix  from the forces of the distorted  molecules \cite{Porezag96}.
The vibrational frequencies were subsequently obtained by diagonalizing the
dynamical matrix. This also allows one to compute the zero-point energies.
We also calculate the IR and Raman spectra of the optimized \cuo{6} molecule which are 
obtained from the derivatives of the dipole moments and polarizability tensors \cite{IR}.

\section{\label{sec:datab}Results and discussion}

 Our test calculations performed for O$_2$ and O$_3$ molecules yield 
the  bond lengths of O$_2$ and O$_3$ as 1.22 and 1.28 \AA which are in excellent 
agreement with the experimental values of 1.21 and 1.28 \AA.  The bond angle 
for O$_3$ is also well predicted, 118$^o$ (calc.) vs. 117$^o$ (expt. \cite{Herzberg2}).
The calculated vibrational frequencies for O$_2$ and O$_3$, 1547 \cmin and  708, 1061, 
1185 \cmin agree well with the experimental values \cite{Herzberg1,Herzberg2} of 
1580 \cmin and 705, 1042, 1110 \cmin respectively. However, the calculated 
dissociation energies of O$_2$ and O$_3$ are 6.80 eV and 9.06 eV while the experimental 
values are 5.1 and 6.2 eV. Although the dissociation energies are grossly overestimated, 
there is good agreement for structural parameters
between the calculated values and experiment.
In the following we discuss the \cuo{x} complexes for x=1,2,3,and 6.

\subsection{CuO}

 The calculated values of bond length, dissociation energy 
and the vibrational frequency of the CuO molecule are given in Table 
\ref{CuO:tab}. Earlier theoretical
values calculated at different levels of theory along with the experimental 
values are also included for comparison.
Our calculated value of bond length 1.72 \AA is in excellent agreement with
experimental value of 1.73 \AA \cite{Polak91}. 
The PBE functional seems to  provide significant improvement over the LSDA and also seem to perform
better than the B3LYP in the bond length prediction.
Our calculated value of the vibrational frequency (680 cm$^{-1}$) is
again in good agreement with experiment (631 cm$^{-1}$).
The dissociation energy is however overestimated in all the
theoretical calculations including the present one. 

\begin{table}
\caption{
The calculated values of bond length  ($R_e$ in \AA), dissociation energy ($D_e$ in eV), 
and the frequency ($\omega$ in cm$^{-1}$) of CuO are compared with the 
experimental values as well as earlier theoretical values. }
 \label{CuO:tab}
\begin{ruledtabular}
\begin{tabular}{lllll}
 $R_e$    &  $D_e$  &    $\omega$ & EA  & Method   \\
 $\AA$    &   eV  &    \cmin & eV & \\
\hline
 
 1.73    &  2.75   &   631    & 1.78    &  EXPT.\footnotemark[1] \\
         &         &   631    &         &  EXPT.\footnotemark[2] \\
 1.78    &         &   640    &         &  EXPT.\footnotemark[3] \\
 1.72    &  3.54   &   680    & 1.84    &  PBE\footnotemark[4] \\
 1.74    &         &   592    &         &  CASSCF\footnotemark[5]\\
 1.73    &  3.58   &   630    &         &  PWPP-GGA\footnotemark[6] \\
 1.68    &  4.13   &   670    &         &  PWPP-LSDA\footnotemark[6] \\
 1.77    &         &   586    &         &  B3LYP\footnotemark[7] \\
 1.69    &  4.01   &   735    & 1.68    &  LSDA \footnotemark[8] \\
\end{tabular}
\end{ruledtabular}
\footnotetext[1]{Experimental results, Ref. \onlinecite{Polak91}}
\footnotetext[2]{Experimental results, Ref. \onlinecite{TMMS81,AL74}}
\footnotetext[3]{Experimental results, Ref. \onlinecite{WDW97}}
\footnotetext[4]{ {Present Calculations} (Zero-point energy is included in $D_e$.)}
\footnotetext[5]{Complete active space self-consistent field method, Ref. \onlinecite{MTK91}}
\footnotetext[6]{Plane wave pseudo-potential method, Ref. \onlinecite{PMC00}}
\footnotetext[7]{Ref. \onlinecite{CAB97}}
\footnotetext[8]{Ref. \onlinecite{WWDL96}}
\end{table}

\subsection{\cuo{2}}

\begin{figure}[b]
\epsfig{file=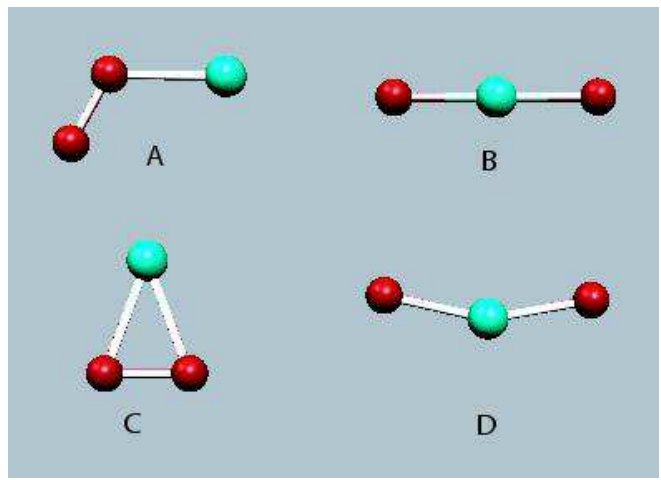,width=\linewidth,clip=true}
\caption{\label{fig:cuo2} The various isomers of the \cuo{2} cluster. The isomer
\AAA is the lowest energy structure. Isomers A, B, and D are vibrationally stable.
}
\end{figure}

   The optimized structures of various isomers of the \cuo{2} molecule are shown 
in Fig. \ref{fig:cuo2}.

\begin{table*}[t]
\caption{
 Calculated bond lengths ($R_e$ in \AA), atomization energies ($D_e$ in eV), 
and vibrational frequencies ($\omega$ in cm$^{-1}$) of \cuo{2}. 
The experimental values as well as earlier theoretical values are included for comparison.}
 \label{OCuO:tab}
\begin{ruledtabular}
\begin{tabular}{llllllllll}
 Structure & Method & \multicolumn{2}{c }{$R_e$}& $\angle$OCuO & $\mu_B$  & $D_e$  & \om{1} & \om{2}  & \om{3}   \\
        &         &       Cu-O   & O-O  &      &       &      &     \cmin      &    \cmin       &   \cmin          \\
\hline

Bent  ({\bf A})    & DFT/PBE\fn[2]     & 1.88  & 1.29 &     & 1 & 2.57 &  240  &   452    &  1200 \\
                   & DFT/B3LYP\fn[6]   & 1.91  & 1.3  &     &   &      &  217     &   438    &  1188 \\
                   & CI\fn[4]          & 1.88  & 1.34 & 120 & 1 &      &          &          &       \\
                   & CI\fn[5]          & 1.91  & 1.34 & 106 & 1 &      &          &          &       \\
                   & CI\fn[10]         &       &      &     & 1 &      &  220     &   535    &  1161 \\
                   & DFT/B3LYP\fn[9]   & 1.92  & 1.3  &     &   &      &  214     &   439    &  1175 \\
                   & CASSCF\fn[7]      &       &      &     &   &      &  96     &   503    &  1359  \\

Linear ({\bf B})   & EXPT.\fn[1]       &       &     &     &   &      & 193   &    658   & 819,823 \\
                   & DFT/PBE\fn[2]     & 1.67  &     &     & 1 & 2.32 & 146   &    705   &   904 \\
                   & DFT/GGA\fn[3]     & 1.68  &     &     &   & 2.27 & 122   &    663   &   864 \\
                   & CI\fn[4]          & 1.77  &     &     & 1 &      &       &              \\
                   & CI\fn[5]          & 1.74  &     &     & 1 &      & 591   &              \\
                   & DFT/B3LYP\fn[6]   & 1.65  &     &     &   &      & 142   &  727     & 953     \\

Side-on  ({\bf C})& DFT/PBE\fn[2]      & 1.97  & 1.39 & 41  & 1 & 2.39  &  287i   &   260    &  940 \\
                  & DFT/B3LYP\fn[6]    &       &      &     &   &       &  171    &  401     & 1036 \\
                  & DFT/B3LYP\fn[9]    & 1.97  & 1.37 & 70  &   &       & 173i   &  453     & 1119 \\
                  & CI\fn[4]           & 2.08  & 1.49 & 42  &   &\\
                  & CASSCF\fn[7]       & 1.99  & 1.86 &     &   &       &  224     &   391    &  1052 \\

                  & EXPT.\fn[6]        &       &     &      &   &      &       &  540,548  & 1089  \\

Wide-angled ({\bf D})& DFT/PBE\fn[2]   & 1.73  & 3.39 & 157 & 3 &2.40 &  175    &   573    &  730 \\
\end{tabular}
\end{ruledtabular}
\footnotetext[1]{Experimental results from Ref. \onlinecite{BE84} and \onlinecite{CAB97}}
\footnotetext[2]{Ref. {Present Calculations}, Zero-point energy is included in $D_e$.}

\footnotetext[3]{Ref. \onlinecite{DYYZ99}}
\footnotetext[4]{Ref. \onlinecite{HN85}}
\footnotetext[5]{Ref. \onlinecite{MT91}}
\footnotetext[6]{Ref. \onlinecite{CAB97}}
\footnotetext[7]{Ref. \onlinecite{BLPS93}}
\footnotetext[9]{Ref. \onlinecite{BA96}}
\footnotetext[10]{Ref. \onlinecite{MNYK89}}
\end{table*}

     The bent CuOO structure ({\bf A} in Fig.\ref{fig:cuo2}) with Cs symmetry is the 
lowest energy structure
amongst the Cu(O$_2$) complexes.  It is  lower than the linear OCuO  molecule 
(structure {\bf B}) by 0.76 eV.
The  third structure with C$_{2v}$ symmetry ( {\bf C} in Fig. \ref{fig:cuo2} ) 
in which the  Cu atom  occupies a side on position with respect to the O$_2$ 
is higher in 
energy by  0.55 eV.  The fourth isomer with C$_{2v}$ symmetry (isomer {\bf D} in Fig. 
\ref{fig:cuo2}) is not reported 
in earlier works and is found to be lower than the structure {\bf C} by 0.02 eV. 
In this wide-angled isomer 
the distance between the two oxygen atoms is much larger than that in the third structure. The OCuO angle in 
{\bf C} is $41^0$ while that in the isomer {\bf D} is $157^0$. Our vibrational
frequency calculations have shown that the side on {\bf C} isomer has one imaginary frequency. 
Such imaginary frequency for this isomer has also been observed in the earlier
density functional based work of Barone and Adamo\cite{BA96}.
 Earlier quantum chemical calculation by Bauschlicher\cite{BLPS93} found isomer {\bf C}
to be more stable than the bent structure {\bf A}. They however note that the two structures are energetically
very close with their order being dependent on the numerical details.
 More recent quantum chemical calculation (CCSD) using the effective core potential (ECP)
reports the side-on  structure {\bf C} to be the lowest. The differences are possibly 
due to the use of ECP.
The bent CuOO structure ({\bf A}) found as the lowest energy structure in the present work
is in agreement with the earlier DFT calculations \cite{PMC00}. However, the order of isomers
as the structures {\bf A, C, and B} in increasing energy observed in earlier
DFT works is now altered in the present work due to the  new isomer ({\bf D}) 
found in this work. This isomer is the next low-lying isomer to the lowest 
energy structure. 
   The calculated values of the  Cu-O and O-O bond lengths, $\angle$OCuO, spin magnetic moment,
atomization energy and vibrational frequencies  are 
compared with the existing theoretical and experimental results in Table \ref{OCuO:tab}. 
The earlier theoretical calculations have predicted the dissociation energy of \cuo{2} with
respect to Cu and O$_2$ to lie between 0.42 -0.56 eV \cite{HN85,BLPS93,BA96}.
The lowest-energy structure  has a doublet state in agreement with the experimental observation 
\cite{BE84, CAB97}.  Our results show that in the linear structure {\bf B}, the Cu-O 
distance is 1.67 \AA which is in good agreement with the prediction based on the experimental  
absorption spectra\cite{Tevault82} and is also consistent with the findings of another 
experimental result\cite{OMG83}.  Another density functional based calculation \cite{DYYZ99} 
using the Perdew-Wang91 GGA \cite{PW91}, however predicts the quartet state to be lower in 
energy by 0.1 eV. Since the two 
GGA functionals, PBE in the present and PW91 in Ref. \cite{DYYZ99}, are essentially
similar, the prediction in Ref. \cite{DYYZ99} could be due to  other numerical
details like choice of  basis set in Ref. \onlinecite{DYYZ99}. They however conclude the
ground state to be a doublet on the basis of vibrational frequencies. 
Density functional calculation by Cherithin \etal \cite{CAB97} using the hybrid 
B3LYP GGA also predicts the quartet to 
be lower than the doublet, but again assign the  doublet to be the ground state on 
the basis of frequency calculations. 
Our predicted values of frequencies are in good agreement with the experimental values
and earlier B3LYP calculation\cite{CAB97}. Unlike previous theoretical calculations,
the present calculation correctly points doublet to be the ground state on the consideration of
both the dissociation energy as well as the frequencies. 

The frequency labeled $\omega_1$ in Table II corresponds to a
bending mode in the bent structure or a flapping mode for the linear and the 
wide-angled structures. The $\omega_2$
refers to a stretching of Cu-O bonds resulting in a breathing mode of vibration
for the symmetric structures. The third frequency corresponds to the O-O stretch mode
in the bent structure while in the linear and wide-angled structures it corresponds to 
the vibration of the Cu atom. The O-O stretching mode of the bent structure at 1200 \cmin
shows a 
large peak in the infrared absorption spectrum while the mode at 240 \cmin is strongly
Raman active. 

\subsection{\cuo{3}}

\begin{figure}[t]
\epsfxsize=3.0in
\epsfbox{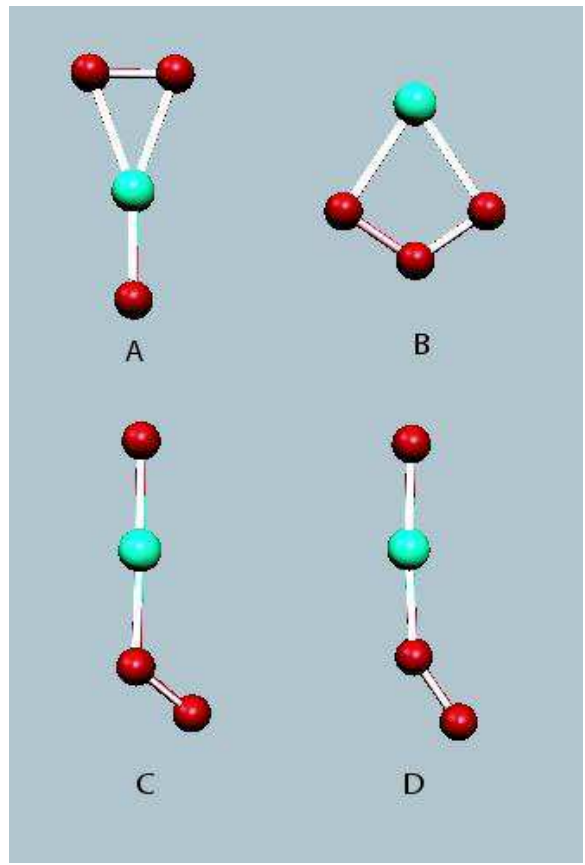}
\caption{\label{fig:cuo3} The various isomers of the \cuo{3} cluster.
}
\end{figure}

 The equilibrium structures of \cuo{3} are presented in Fig. \ref{fig:cuo3}.
 The first structure ({\bf A} in Fig. \ref{fig:cuo3}) has a \YY shape with  Cu 
atom occupying the center of the fork position.  The isomer {\bf B} has a distorted
rhombus structure while {\bf C} and {\bf D} structures are chain-like. One of
the chain structures has a linear O-Cu-O unit while the other one
has a wide-angled O-Cu-O unit.
The spin magnetic moments, dissociation energies and the vibrational frequencies of 
the four isomers 
are given in Table \ref{cuo3:tab}. Our results indicate that the \YY shaped 
cluster containing a side-on \cuo{2} unit 
is the lowest energy structure. The spin state of this structure is  quartet.
This result is in agreement with earlier published DFT calculations \cite{PM02,DYZ00}.
The distorted rhombus structure has the Cu atom bonded with an O$_3$ unit which we refer to as 
copper ozonide structure. This isomer is 
higher in energy by  0.54 eV. The chain-like structures have energies lying between the
\YY shaped structure and the ozonide structure. 

\begin{table}
\caption{The spin magnetic moments (in bohr magneton), atomization energies (eV) and vibrational frequencies of the
various isomers of \cuo{3}. }
\label{cuo3:tab}
\begin{ruledtabular}
\begin{tabular}{lllll}
 Isomer   & Method  &   $\mu_B$& $D_e$  & \om{e} \cmin  \\
\hline
 
   & EXPT.   \fn[1]  &        &        &           660(90)  \\
A  & DFT/PBE\fn[3]   &    3   & 2.92   &  136, 143, 273, 378, 703, 1158 \\
   & DFT/PW91\fn[4]  &    3   & 2.90   &           366,683,1160 \\
   & DFT/B3LYP\fn[2] &    3   &        &           386, 656, 1230 \\

B  &  DFT/PBE\fn[3]  &    1   & 2.79   & 195,229,321,698,807,987  \\
   &  DFT/PW91\fn[4] &    1   &        &             697, 799, 977 \\
   &  DFT/B3LYP\fn[2]&    1   &        &             724, 879, 1055 \\
   &  EXPT.   \fn[2] &        &        &  792, 802, 812     \\

C  &  DFT/PBE\fn[3]  &    3   & 2.87   &  69, 95, 206, 458, 676, 1267 \\
   &  DFT/PW91\fn[4] &    3   &        &               441, 654, 1258 \\
   &  DFT/B3LYP\fn[2]&    3   &        &               400, 635, 1351 \\

D  &  DFT/PBE\fn[3]  &    1   &  &  98, 163, 277, 468, 800, 1358  \\

\end{tabular}
\end{ruledtabular}
\footnotetext[1]{Experimental results, Ref. \onlinecite{WDW97}.}
\footnotetext[2]{Ref. \onlinecite{CAB97}.}
\footnotetext[3]{Ref. {Present Calculations}, Zero-point energy is included in $D_e$.}
\footnotetext[4]{Ref. \onlinecite{DYZ00}.}
\end{table}

   The calculated vibrational frequencies of all the four isomers are real indicating that all the 
four structures
are local minima on the potential energy surface (Cf. Table \ref{cuo3:tab}).  
 In case of the lowest energy structure (Structure {\bf A}), the lowest two modes correspond 
to bending motions, the first one in plane
and the second one perpendicular to the plane of the molecule. The third mode pertains
to the rotational motion of the O$_2$ unit about an axis perpendicular to the 
O-O bond.  The next two modes correspond to Cu-O bond stretch modes. 
The largest frequency (1158 \cmin) mode refers to the O-O stretch mode.  Similar to the 
smaller clusters, this mode of vibration is strongly IR active while Cu-O stretch modes show 
significant Raman scattering.

 In the ozonide isomer ({\bf B}), the lowest three frequencies respectively involve the asymmetric 
stretch of the Cu-O
bonds, vibration of the central O perpendicular to the plane of the molecule and a
symmetric stretch of the Cu-O bond leading to a breathing mode.
The frequencies from 687 to 987 \cmin pertain to the flapping , asymmetric O-O stretch and a symmetric
O-O stretch modes of the ozone unit. However, the frequencies are lower than a free
ozone unit indicating weaker O-O bonds in the O$_3$ unit due to the  presence of the
Cu atom. The mode at 807 \cmin shows strong infrared absorption while those around 200 \cmin
are highly Raman active. Cherithin et al. \cite{CAB97} have observed peaks around 800 \cmin
(812, 802 and 792 \cmin) in their IR measurement which they have attributed to the
Cu-ozonide structure. The present results are supportive of this assignment.

 In the two chain-like structures the vibrational frequencies are comparable and the
modes are same. The lowest two modes correspond to the vibration of the Cu atom
in and out of the plane of the molecule. The next two frequencies refer to the asymmetric
Cu-O stretch modes while the largest frequency corresponds to the O-O stretch mode
which is also strongly IR active. The vibrational frequency of the O-O unit
in the linear modes tends towards that of the free O$_2$ frequency (1547 \cmin). 
The Cu-O stretch modes display Raman activity.  The dissociation energies of the 
chain-like molecules {\bf C} and {\bf D} into CuO and O$_2$ are 1.14 and 0.76 eV.  

\subsection{\cuo{6}}

        In the CuO$_3$ clusters, we have seen that the ozonide
cluster is higher in energy than  the \YY shaped structure which contains an O$_2$
unit. On the other hand, the CuO$_2$ triangular structure is higher in energy than
a linear Cu-O-O structure. It would be interesting to see how the cluster structure 
changes in the presence of more oxygens.

\begin{figure}[b]
\epsfig{file=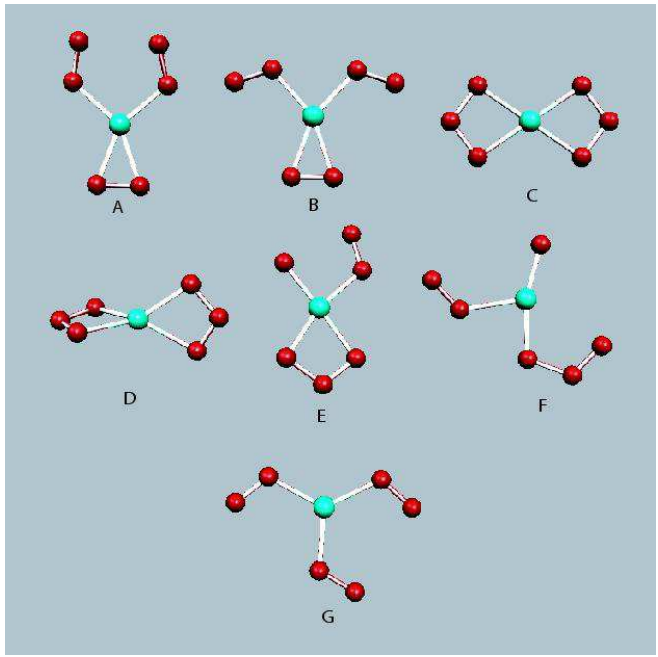,width=\linewidth,clip=true}
\caption{\label{fig:cuo6} The various isomers of the \cuo{6} cluster. The isomer A is 
the lowest-energy isomer.
}
\end{figure}

        The knowledge of structures for the smaller sizes is useful in building
up the starting geometries for the optimization of \cuo{6}. We have started with 
several possible structures as input for optimization. The optimized geometries 
are presented in Fig.~\ref{fig:cuo6}. 
The first two structures, labeled as {\bf A} and {\bf B} in the Fig.\ref{fig:cuo6}, 
contain the three O$_2$ units and also a CuO$_2$ triangle.
The difference in the structures lie in the orientations of the open-ended O$_2$
units. The  isomers {\bf C} and {\bf D} contain closed Cu-O$_3$ rings. 
The isomer {\bf C} is planar and the isomer {\bf D} is 3-dimensional. The structure 
{\bf E} has both O$_3$ and  O$_2$  units of which the O$_3$ forms a closed 
ring with the Cu atom.  For the  next isomer (\FF) also, both O$_3$ and O$_2$ units are 
present and the Cu atom forms  end-on bonds with these units.
The last structure \GG has three O$_2$ units 
all of which form end-on bonds with the Cu atom. This structure was arrived
at by starting the geometry optimization with the structure predicted by Wang et al.
based on their photoelectron spectra. Our calculations have shown the
structure labeled as {\bf A} has the lowest energy among all the structures shown here.
The structures {\bf B} and {\bf G} are comparable in energy and both are higher
relative to structure {\bf A} by 0.12 and 0.22 eV respectively.
The isomers \CC and \DD with ozonide subunits  are higher 
than {\bf A} by 0.95 and 0.94 eV respectively.
The ozonide planar structure {\bf C}  was reported to be  the lowest energy structure by
Pouillon and Massobrio \cite{PM02}. This difference in the present and their 
calculations for the lowest energy structure is probably due to their choice
 of ultra-soft pseudopotential for representation 
of core electrons and brings out the importance of treating all electrons at equal footing
in the CuO$_x$ system.  The structures {\bf E} and {\bf F} are higher in energy by 
1.62 and 1.79 eV, respectively.
The atomization energies of these isomers are presented in Table \ref{cuo6:tab}.
The atomization energy includes the zero-point energy of the systems.

The vibrational   frequencies of the isomers are presented in Table \ref{cuo6:tab}.
Both the structures {\bf A} and {\bf B} are found to be vibrationally stable. 
In these two lowest energy structures {\bf A} and \BB, the modes of vibrations are pretty
similar. The low frequency modes are associated with the bending and flapping of
the open-ended O$_2$ units. On the other hand the high frequency modes are associated
with the stretching of the O-O bonds reflecting the strength of these bonds. The
Cu-O bond stretch modes have frequencies lying between 250 to 525 \cmin. The lower
frequencies mainly correspond to the bond bending motions of the O$_2$ units. In the
structure {\bf A} the high frequency modes corresponding to the O-O bond stretching
show significant IR absorption while the modes below 400 \cmin show Raman activity.
On the other hand, the structure {\bf B} is IR inactive and has modes below 200 \cmin 
that are Raman active.

 The structure \CC comprising an ozonide unit is also vibrationally stable. This structure 
is both Raman and IR active. The Raman scattering is seen in the low frequency range of 300-350 
\cmin, where the modes correspond to bending of the molecule and also symmetric and 
asymmetric stretching of the four Cu-O bonds. The IR active modes are the high frequency 
modes in the range of 800 - 920 \cmin. These are again O-O stretch modes which are asymmetric.
The largest mode is also a symmetric O-O stretch mode leading to a breathing mode. However, 
unlike the other structure where the largest mode is strongly IR active, this mode is IR inactive.
This is due to the fact that this mode is symmetric and  does not lead to a change in 
dipole moment of the molecule.
\begin{table*}[t]
\caption{The atomization energies, spin states and the vibrational frequencies of the \cuo{6} isomers.
The dissociation energies  include zero-point energy.}
 \label{cuo6:tab}
\begin{ruledtabular}
\begin{tabular}{llllc}
 Isomer   & $D_e$ (eV)& $\mu_B$  & \om{e} \cmin  \\
\hline
 
  A &   3.29  & 3   & 30, 95, 107, 128, 148, 174, 264, 294, 299, 405, 476,
496, 1147, 1250, 1324  \\

 B  & 3.27  & 1    & 36, 69, 93, 101, 124, 141, 192, 236, 359, 381, 439, 525, 1168, 1233, 1298 \\

  C &  3.15  & 1    & 70, 94, 116, 311, 312, 347, 365, 437, 456, 706, 722, 810, 826, 916, 1015  \\


 D &   3.16  & 1   &  64$i$, 49, 88, 167, 246, 295, 342, 364, 389, 667, 751, 789, 909, 912, 1030 \\


 E &  3.06  &  3     &  53, 84, 103, 157, 170, 232, 303, 387, 494, 636, 695, 790, 993, 1340 \\

 F &  3.03&  1      & 29, 74, 87, 104, 128, 155, 183, 299, 364, 421, 570, 703, 752, 1179, 1400 \\


  G &   3.26  &  5  &  25i,44,45,78, 94, 104, 175, 184, 221,  373, 387, 398, 1240, 1251, 1323 \\

\end{tabular}
\end{ruledtabular}
\end{table*}

 The  structure {\bf D} with two orthogonal planar O$_3$ units, is a saddle point with an  
imaginary vibrational frequency.  The eigenvector of the mode shows it related to a bending 
motion which try to break the orthogonality of the two ozone units. This orthogonality was 
not enforced during the optimization.  Both the structures {\bf E} and {\bf F} containing 
both O$_2$ and O$_3$ units are found to be vibrationally stable. However, they are very high 
on the energy scale. This is again consistent with the energy ordering seen in the \cuo{3} 
isomers in which both the ring-like and linear structures are higher in energy than the \YY 
shaped one.  The vibrational analysis of the structure \GG showed it to be a saddle point on 
the potential energy surface.  The \GG isomer containing three open-ended O$_2$ units shows 
IR activity in the range of 1200-1250 \cmin.  These again are O-O stretching modes. However 
the largest frequency mode which corresponds to symmetric stretching of all the O-O bonds is 
found to be IR inactive.  This can be attributed to the fact that this mode retains the overall 
symmetry of the molecule and does not lead to a change of the dipole moment. 
\begin{table*}[t]
\caption{The magnetic moment M (in Bohr magneton), relative energies $\Delta E$,  vertical
detachment energy (VDE) and the vibrational frequencies of the \cuo{6} anionic isomers.}
\label{anion:tab}
\begin{ruledtabular}
\begin{tabular}{llllc}
 Isomer   &    M ($\mu_B$) & $\Delta E$ (eV)& VDE (eV)  & \om{e} \cmin  \\
\hline
 
A &    2 & 0.00 &  3.12  & 51, 71, 124, 130, 158, 185, 248, 265, 398, 413, 491, 499, 1138, 1209 \\

B   &   2 & 0.06 &3.35 & 27,103, 131, 179, 185, 226, 249, 257, 417, 447, 489, 546, 907, 1133, 1201 \\

C  &   2 & 0.53& 3.25 & 58, 101, 127, 261, 286, 315, 324, 413, 458, 647, 669, 729, 749, 831, 939  \\

D    &   2 & 0.49 & 3.33 & 29,79,80, 233, 234, 238, 291, 291, 397, 665, 794, 794, 932, 946    \\


E  &   2 & 0.89 & 3.49 &  20, 37, 62, 80, 105, 114, 172, 220, 232, 410, 507, 681, 707, 1143, 1299 \\

G  &   4 & 0.36 & 2.80 & 63i, 31i, 51, 65, 105, 130, 166, 196, 204, 379, 390, 405, 1093, 1105 \\
   
\end{tabular}
\end{ruledtabular}
\end{table*}

\begin{figure}[b]
\epsfig{file=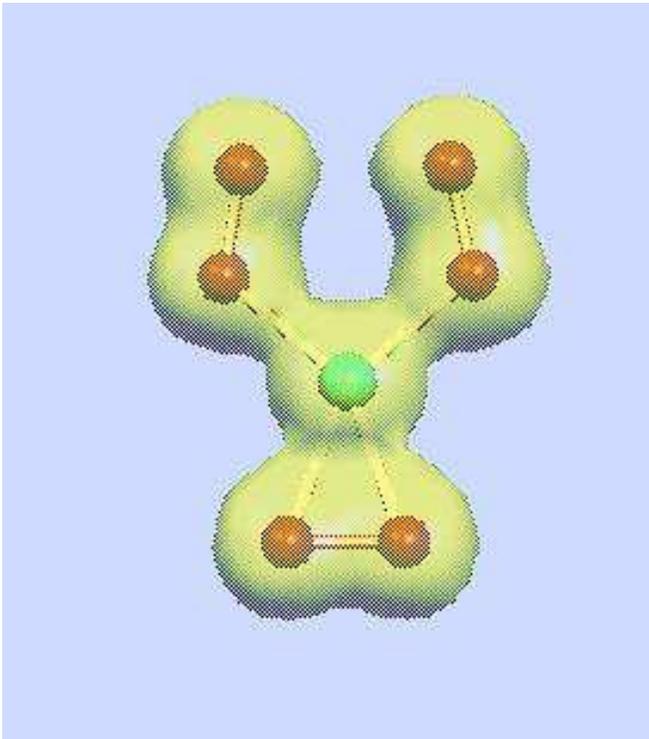,width=\linewidth,clip=true}
\caption{\label{fig:density} The isodensity surface of the lowest energy \cuo{6}
structure.  }
\end{figure}

  We have also considered various possible anionic structures of the \cuo{6} molecule. 
The optimized structures of the anion are very similar to their neutral counterparts 
and are therefore not shown.  These structures are obtained by starting the 
optimization with the geometries of neutral clusters.  As in the neutral case, the 
structure {\bf A}  is found to be the lowest-energy structure.  The relative energies, 
vibrational frequencies, vertical detachment energies of these structures  are given 
in Table \ref{anion:tab}.  An analysis of the vibrational frequencies show that the 
structures {\bf A, B, C} and {\bf D} pertain to  local minima  on the potential energy 
surface. The structure {\bf E} is the only structure to become somewhat modified from 
its neutral geometry in that the oxygen dimer tilts away from the Cu and Cu-O$_2$ bond 
length increases to 2.3 \AA. Also the CuO$_3$ closed ring structure becomes opened at 
one end.  This structure is vibrationally stable but energetically high.  On the other 
hand, the high energy structure {\bf G} has two imaginary frequencies associated with it. 
The optimization of the anionic counterpart of structure {\bf F} was abandoned 
due to failure to obtain a convergent solution. The structure D in neutral form is 
vibrationally unstable but the anionic form is vibrationally stable.  The adiabatic 
electron affinity of the lowest energy structure is found to be 2.75 eV in which the 
zero point motion of the molecule is also taken into consideration.  The calculated 
values of the VDE (Cf. Table \ref{anion:tab}) are in the range $2.8-3.5$ eV.  The 
experimental value of the electron binding energy in \cuo{6} as measured from 
photoelectron spectra is 3.2 eV \cite{WDW97}.  The structures with ozonide units as well 
as with O$_2$ units, both have the VDE around this value and based on the consideration 
of the VDE alone, it is difficult to infer about the correct anion geometry. However, the 
VDE combined with total energy values indicate the oxide structure to be the candidate for 
the ground state geometry for the neutral \cuo{6}.
  
    The analysis of nature of bonding based on the charge density differences
shows the expected behavior, that the O-O bonding is covalent while Cu-O bonds 
are more ionic in nature. The plot of isosurface of charge density are shown in 
Fig. \ref{fig:density} for the case \cuo{6}.  The lowest energy \cuo{6} is stable by 2.72 eV 
against dissociation into Cu atom and three O$_2$ molecules. This translates to 
about 0.68 eV per Cu-O bond which is a relatively weak bond. A simple analysis of 
integrating charge density around each atom inside an atomic sphere of arbitrary 
radius shows the Cu atom to be in the +1 charge state and the O atoms
in the -2 charge state. There is very little spin density on the Cu atom and most
of the spin density is seen on the O$_2$  units. This is consistent with the paramagnetic
nature of O$_2$ molecule. The spin density is largest on the O atoms farthest from
the Cu atom.

\section{\label{sec:summary}{Summary}}

 In conclusion, we have performed a detail examination of various possible
structures of the CuO$_x$ molecule $x=1-3,6$. The vibrationally stability of the isomers 
was examined by carrying out an analysis of the vibrational modes.  A bent structure 
with C$_s$ symmetry is found to be the most stable structure for \cuo{2}. On the other hand, 
a \YY shaped structure with C$_{2v}$ symmetry is the lowest energy structure for \cuo{3}.  
Our study shows the potential energy surface is rather flat for \cuo{6}.
Within our fairly exhaustive search we find two energetically nearly degenerate 
isomers as possible candidate structures for the ground state. Both of these
structures contain \cuo{2} subunits. This observation is consistent with 
the argument by Wang and coworkers based on the PES but it rules out their proposal of the
structural form. Further, contrary to the earlier pseudopotential based study, 
the structures with ozonide subunits are found to be  high energy isomers.
The anionic clusters show trends similar to those of their neutral counterparts.
The \cuo{6} cluster in its most stable form is found to be strongly IR active. 
The IR active modes correspond to the O-O bond stretching and  generally have high frequencies.

\section{Acknowledgment}
  TB acknowledges helpful discussion with S. L. Richardson.
  TB and MRP acknowledges financial assistance from ONR (Grant No. N000140211046) and
by the DoD High Performance Computing CHSSI Program. RRZ acknowledges support from the GMU.

\end{document}